# Interference Effects of the Superconducting Pairing Wave Function due to the Fulde-Ferrell-Larkin-Ovchinnikov like State in Ferromagnet/Superconductor Bilayers

V. I. Zdravkov[1,2], J. Kehrle[1], G. Obermeier[1], A. Ullrich[1], S. Gsell[1], D. Lenk[1], C. Müller[1], R. Morari[2], A. S. Sidorenko[2], V. V. Ryazanov[3], L. R. Tagirov[1,4], R. Tidecks[1], and S. Horn[1]

[1]*Institut für Physik, Universität Augsburg, D-86159 Augsburg, Germany*
[2]*Institute of Electronic Engineering and Nanotechnologies ASM, MD2028 Kishinev, Moldova*
[3]*Institute of Solid State Physics, Russian Academy of Sciences, 132432 Chernogolovka, Russia*
[4]*Solid State Physics Department, Kazan State University, 420008 Kazan, Russia*

**Abstract**

The theoretical description of the Fulde-Ferrell-Larkin-Ovchinnikov like state establishing in nanostructered bilayers of ferromagnetic (F) and superconducting (S) material leads to critical temperature oscillations and reentrant superconductivity as the F-layer thickness gradually increases. The experimental realization of these phenomena is an important prerequisite for the fabrication of the Ferromagnet/Superconductor/Ferromagnet core structure of the superconducting spin-valve. A switching of the spin-valve is only expected if such non-monotonic critical temperature behavior is observed in F/S bilayers as well as in the S/F bilayers, a combination of which the spin-valve core structure can be regarded to consist of. In our former investigations we could demonstrate the required non-monotonic behavior of the critical temperature in S/F bilayers. In this study we succeeded in the preparation of F/S bilayers, where the superconducting material is now grown on top of the ferromagnetic metal, which show deep critical temperature oscillations as a function of the ferromagnetic layer thickness as well as an extinction and recovery, *i.e.* a reentrant behavior, of superconductivity. Especially, the latter is necessary to obtain a spin-valve with a large critical temperature shift between the parallel and antiparallel configurations of magnetizations in the F layers.

## 1. Introduction

Conventional singlet superconductivity and ferromagnetism are antagonistic long-range orders. Aside a vanishing total momentum, superconductivity implies zero total spin of a Cooper pair, whereas ferromagnetism imposes parallel arrangements of the conduction electron spins, thus destroying singlet superconductivity. A way to resist against the destructive alignment of the conduction electron spins in a homogeneous material was proposed by Fulde-Ferrell and Larkin-Ovchinnikov (FFLO) [1,2]. In their theory the Cooper pair still has a zero total spin, but acquires a non-vanishing pairing momentum. The range of parameters for which the FFLO state can be observed in a bulk material is extreme and narrow [3] and, thus, hard to be realized [4-7]. An induced FFLO state can, however, be generated in nanolayered thin-film structures by the proximity of superconducting (S) and ferromagnetic (F) metals [8]. Contrary to the conventional S/N proximity effect, in which the pairing wave function decays exponentially into the normal conducting non-magnetic material (N), in the S/F proximity effect this decay is modulated by oscillations caused by the non-vanishing pairing momentum (see Refs. [8-10]). If the magnetization of the F-layer is inhomogeneous (e.g., if it is in the magnetic domain state), moreover, triplet components of the superconducting pairing can be generated [10,11].

Due to the oscillation of the pairing wave function, interference effects occur in S/F layered structures similar to the optical case of light in a Fabry-Pérot interferometer, yielding an oscillation of the critical temperature, $T_c$, with increasing thickness of the F layer. For a suitable set of parameters even a complete extinction and recover, *i.e.* a reentrant behavior of the superconducting state may occur, as predicted theoretically in Ref. [12]. Recently, deep $T_c$ oscillations and the first convincing experimental realization of the reentrant behavior could be demonstrated in niobium/copper-nickel-alloy bilayers [13] and was subsequently studied in detail in this couple [14]. Also, evidence for the multi-reentrant state, predicted by theory [12] was found during these investigations [14].

The strong and phase-dependent impact of magnetism on superconductivity provoked several ideas how to use ferromagnet-superconductor heterostructures to build functional superconducting devices. The most elaborated one is the so-called π-junction [15,16], where the intrinsic FFLO-phase changes by π across the ferromagnetic weak link in an S/F/S Josephson junction. It was exploited to fabricate π-phase-shifters for superconducting digital and quantum circuits (see, for example, [17,18]).

Another possible application is to control the transport supercurrent making use of an F-electrode/Superconductor/F-electrode spin valve [19,20]. Although a series of attempts to fabricate such type of spin-valve have been undertaken recently [21-37], the observed effect of variations of the superconducting transition temperature, $T_c$, upon changing the alignment of magnetizations in the electrodes from parallel to antiparallel was very tiny (from several to some dozens of milli-Kelvin). The reason is probably a non-optimal choice of materials and layer thicknesses.

It has been shown theoretically ([19], Fig. 3; [14], Fig. 8, where the spin-valve was called "spin-switch") that a reentrant behavior of superconductivity in an F/Superconductor/F trilayer at parallel alignment of magnetizations would be optimal to get a much larger effect (in the Kelvin range, if calculated with the model parameters estimated for the Nb/Cu$_{41}$Ni$_{59}$ couple [14]).

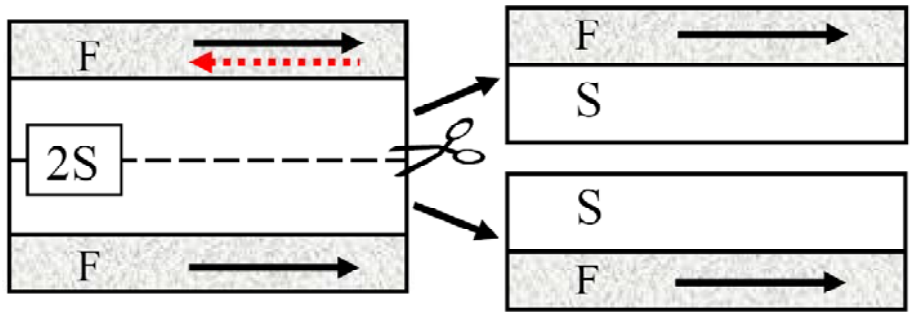

**Fig. 1.** F/2S/F trilayer (left-hand side) as a stack of two bilayers (right-hand side): F/S and S/F. If the magnetizations (ARROWS) in the F/2S/F trilayer are aligned parallel (left side, BLACK – BLACK), or antiparallel (left-side, dashed-RED – BLACK), the interference conditions for the phase-dependent wave-function change, and the zero spin-projection triplet component of pairing is generated at antiparallel alignment, providing the channel to control superconductivity in the S layer [19,38,39]. If the trilayer stack is split along the symmetry plane (left-side, dashed black line) into two decoupled bilayers (right-hand side), the magnetization direction has no longer an influence on the superconductivity.

If we split the trilayer into two bilayers as it is shown in Fig. 1, the superconducting properties of each of the bilayers should be equal to that of the trilayer because of symmetry considerations. In reality, the growth conditions for the bottom F layer on a substrate (or an exchange-bias layer) and for the top F layer on the surface of the S layer are different. The same is the case for the S layer grown on an F metal and a substrate, respectively. As a result, the specular symmetry of the physical properties for the bilayers in Fig. 1 breaks down. Being combined backward to get the F/2S/F trilayer the F layers may act on superconductivity incoherently, thus diminishing the spin-switch effect.

In our previous work we studied superconducting properties of niobium/copper-nickel-alloy bilayers deposited in the sequence substrate/Superconductor/Ferromagnet, *i.e.* S/F bilayers [13,14]. A special custom setup with moving magnetron to deposit ultra-flat superconducting films up to 8 cm in lateral extent, and a special wedge technique to produce a series of up to 40 samples at the same run was necessary to obtain the deep $T_c$ oscillations and the reentrant behavior of superconductivity as a function of the F layer thickness, mentioned above. The S/F bilayers may be regarded as the first building block of the superconducting spin valve. The next non-trivial problem on the way to the fabrication of the spin valve is the deposition of high quality Ferromagnet/Superconductor bilayers, where now the ferromagnet has to be grown on the Si substrate and the superconductor on top of the F material. The results on this second building block of superconducting spin valve will be reported in the present paper.

## 2. Sample Preparation and Characterization

### 2.1 Thin Film Deposition and Sample Configuration

The samples were prepared by magnetron sputtering on commercial (111) silicon substrates (80×7 mm$^2$ size) at room temperature. The base pressure in the "Leybold Z400" vacuum system was about $2\times10^{-6}$ mbar. Three targets, Si, Nb, and $Cu_{40}Ni_{60}$ (75 mm in diameter) were used, and pure argon (99.999%, "Messer Griesheim") as sputter gas. The targets were pre-sputtered for 10-15 minutes in the parking position to remove contaminations. Moreover, Nb acts as a getter material, to reduce the residual gas present in the chamber.

To fabricate a series of F/S samples with variable layer thickness at the same run, so that the deposition conditions for all samples in the series are identical, we applied our wedge technique described in detail in Ref. [13,14,40]. A wedge of ferromagnetic material ($Cu_{41}Ni_{59}$ alloy) was grown on a silicon buffer layer which was deposited first after the pre-sputtering procedure. The wedge is obtained utilizing the off-symmetry mounting of the long substrate and the intrinsic spatial gradient of the deposition rate of the magnetron sputtering setup [40]. RF sputtering is applied to keep the composition of the deposited layer (41/59) close to the composition of the $Cu_{40}Ni_{60}$ target. The growth rate of the film directly under the magnetron is about 3-4 nm/s.

Immediately after, an ultra flat superconducting niobium layer of constant thickness is deposited, making use of our custom, rotating target technique ("spray" technique), described in detail in Refs. [13,14,40,41]. During uniform movement of the magnetron above the substrate, the Nb target was DC sputtered with the effective growth rate of about 1.3 nm/s. The deposition rate for a fixed, non-moving target would be about 4-5 nm/s. Afterwards, the $Cu_{41}Ni_{59}$-wedge/Nb bilayer was coated by a thin amorphous Si cap-layer to prevent a degradation at ambient conditions. A sketch of the wedge is shown in the inset of Fig. 1b.

Finally, samples of equal width (about 2.5 mm) were cut from the wedge perpendicular to the thickness gradient, resulting in a batch of F/S bilayer strips with varying F layer thickness, $d_F$, for $T_c(d_F)$ measurements. For four-probe resistance measurements, with alternating polarity to eliminate thermoelectric effects, Al wires (50 μm thick) were attached by ultrasonic bonding.

To get samples for $T_c(d_S)$ measurements, the same procedure was applied, however, now a wedge-shaped Nb layer was grown on a $Cu_{41}Ni_{59}$ film of constant thickness.

The special challenge is that, contrary to our previous work, the superconducting material is now grown on the ferromagnetic alloy deposited on the Si substrate with a buffer layer. Since this means completely different growth conditions, a detailed analysis of the resulting specimens has been done, especially by cross-sectional electron microscopy.

**2.2 Thickness and Composition Analysis**

To determine the thickness of the $Cu_{1-x}Ni_x$ and Nb nanolayers of the samples, Rutherford Backscattering Spectrometry (RBS) was applied. Furthermore, with this method it is possible to identify the atomic concentration, $x$, of the copper-nickel alloy layer of the sample. The applicability of this method for ultra-thin film specimens has been demonstrated in our previous works [13,14,40].

The measurements were carried out using $He^{++}$ ions accelerated to energy of 3.5 MeV by a tandem accelerator. The backscattered $He^{++}$ ions were detected under an angle of 170° compared to the incident beam. The samples were tilted 7° azimuthally in order to prevent channeling effects in the silicon substrate.

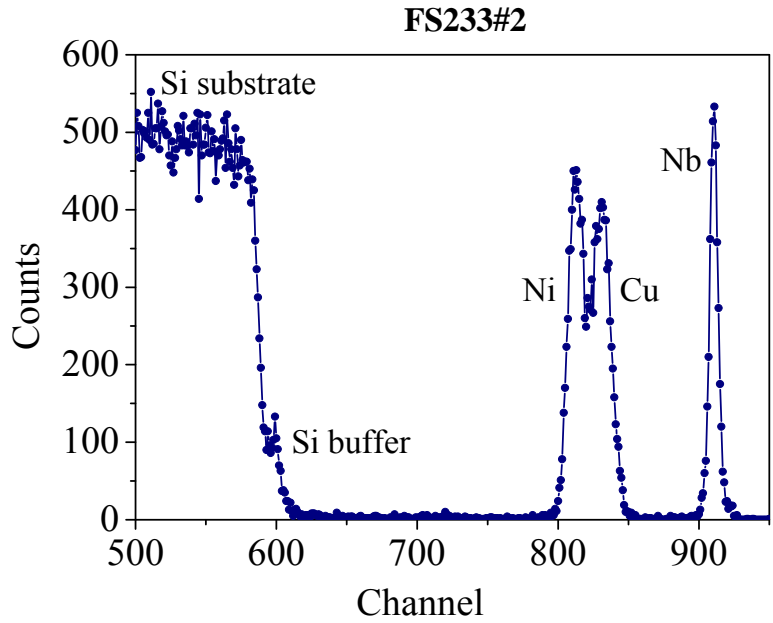

(a)

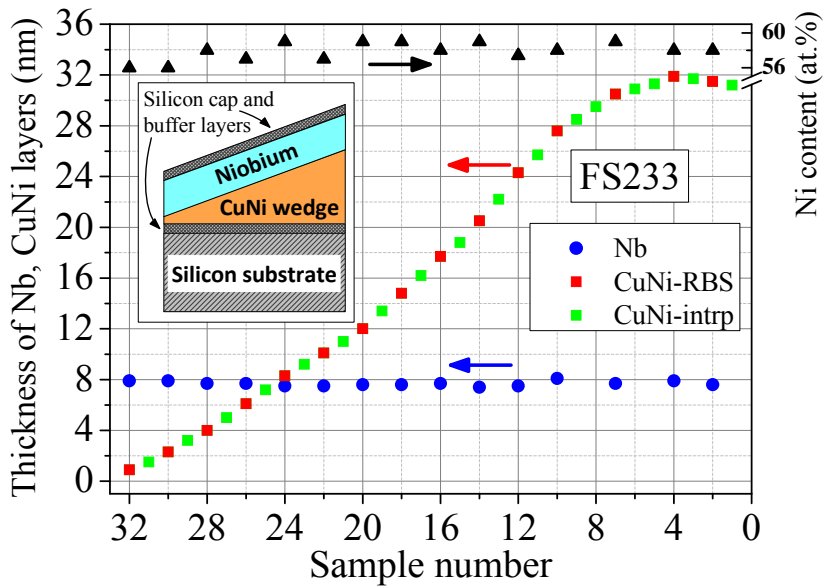

(b)

**Fig. 2.** Rutherford Backscattering Spectrometry: a) RBS spectrum for a $Cu_{1-x}Ni_x$/Nb thin film bilayer on Si substrate; b) Results of the measurements for the thickness of the $Cu_{1-x}Ni_x$ and Nb layers, respectively, together with the nickel content, $x$. The inset shows a sketch of the layers stack. Blue solid dots (Nb), black triangles (Ni content) and red square symbols for the alloy layer thickness are measured points, whereas gray (green in colour) symbols are linear interpolated values.

A typical RBS spectrum is shown in Fig. 2a. By a theoretical simulation, using for instance the RUMP computer program [42], the position of the peaks of the different materials can be identified, and their elemental areal densities can be determined. From the latter the thickness of the Nb and $Cu_{1-x}Ni_x$ layers has been calculated as well as the atomic composition, $x$, of the alloy using the densities of the respective metals.

The results of such evaluation for series FS233 is shown in Fig. 2b. The thickness of the niobium film is nearly constant at a value of $d_{Nb}$(FS 233) = 7.5 nm (-0.4 nm +0.3 nm). The $Cu_{1-x}Ni_x$ layer thickness decreases from 32 nm to 1 nm. The nickel concentration in the $Cu_{1-x}Ni_x$ alloy varies by +1 at.% to -2 at.% around the average value of 58 at.% with a slight decrease of nickel content towards the thin end of the wedge.

## 2.3 Transmission Electron Microscopy

To get direct information about the growth and the thickness of our $Cu_{1-x}Ni_x$/Nb thin-film bilayers, Transmission Electron Microscopy (TEM) images were taken. Fig. 3 a-c shows the cross section of a sample (FS2103 No. 1), obtained by using a JEOL JEM 2100F microscope, equipped with a GATAN Imaging Filter and a CCD camera. The individual layers can be clearly distinguished. Beginning from the silicon substrate at the bottom they are (see Fig. 3a): Si(substrate)/Si(buffer)/$Cu_{1-x}Ni_x$/Nb/Si(cap). The thickness of the copper nickel and niobium layer is 25.4 nm and 8.6 nm, respectively.

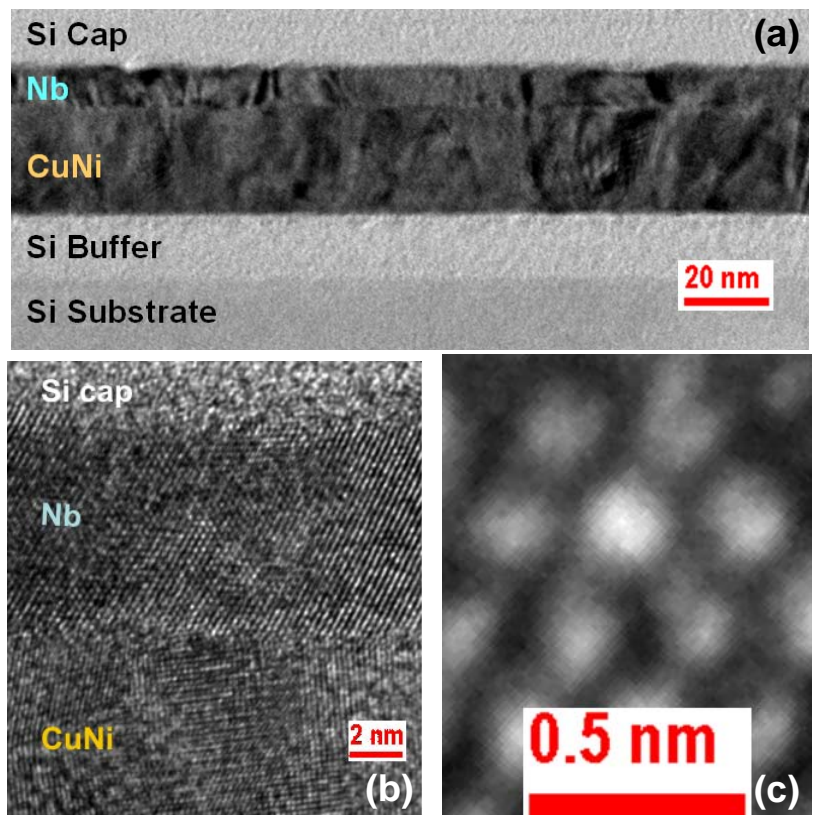

**Fig. 3.** a) TEM image of the layers stack (FS2103 sample No. 1); b) HRTEM view on the Nb and CuNi layers for the same sample, c) three fold symmetry of the Nb (110) lattice planes.

Figure 3b shows a High Resolution TEM (HRTEM) image of the F/S interface. No interface layer has been built up. The small lateral scale roughness of the $Cu_{1-x}Ni_x$ and Nb interface is of atomic scale. Furthermore, the interface is straight without any irregularities like pits or bumps. Also the interface between the silicon cap and the niobium shows a high linearity and sharpness. This is also true for the interface layer build by the copper nickel alloy and the silicon buffer layer.

The spacing between the lattice planes in the HRTEM picture of Fig. 3b is 2.36 Å, which is close to 2.33 Å, expected for the {110} planes in the bcc niobium lattice [43]. This fact and the three fold symmetry seen in Fig. 3c indicates that one is viewing into the <111> direction. Since one of {110} lattice planes (see the horizontal rows of atoms in Fig. 3c) is parallel to the silicon substrate and buffer layer, respectively, the growth of the niobium film occurs with an {110} plane parallel to the substrate.

This growth direction is in agreement with the growth direction of the niobium layer obtained by X-ray diffraction (XRD) in our previous work on niobium/nickel bilayers [40] and in accordance with our XRD measurements on S/F bilayers of $Nb/Cu_{41}Ni_{59}$.

Depending on the area of the $Cu_{41}Ni_{59}$ layer investigated, the lattice planes are separated by a distance of 2.07 Å and 1.85 Å corresponding to the (111) and (200) planes, respectively, of the nickel lattice which is identical to the copper lattice [43]. From this image it is not possible to identify definitely the growth direction of the film.

## 3. Results and Discussion

### 3.1 Superconducting Properties

The resistance measurements of the S/F bilayers were performed in a He-3 cryostat and a dilution refrigerator, using the standard DC four probe method (measuring current 10 µA in the range 0.4-10 K and 2 µA for 40 mK-1 K). Alternating the polarity of the current during the resistance measurements serves to eliminate the thermoelectric voltages. The critical temperature, $T_c$, of the superconducting transition was evaluated from the midpoint of the $R$(T) curves (see Fig. 4). The transition width ($0.1R_N$-$0.9R_N$ criteria, where $R_N$ is the normal state resistance just above $T_c$) for all investigated samples was typically around 0.2-0.3 K.

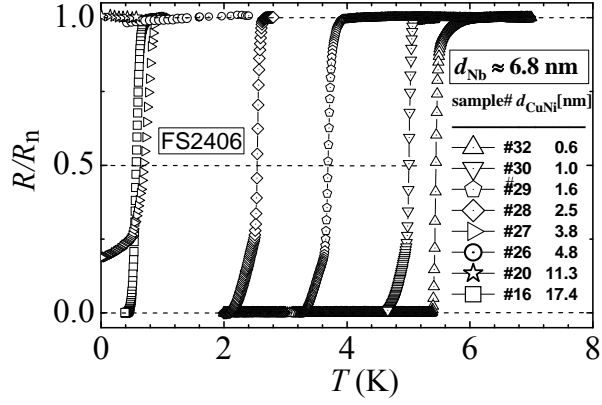

**Fig. 4.** Typical resistive transitions $R(T)$ normalized to $R_n = R(T = 10\ K)$ of the investigated samples.

In Fig. 5a the dependence of the superconducting transition temperature on the thickness $d_{CuNi}$ of the ferromagnetic alloy is shown for different fixed thicknesses $d_{Nb}$ of the superconducting layer. For $d_{Nb} = 7.5$ nm (sample series FS233) the critical temperature $T_c$ shows a non-monotonic behavior, beginning with a decrease, then a minimum is reached, and subsequently the critical temperature rises again.

If the thickness of the flat superconducting layer is further decreased to $d_{Nb} = 6.8$ nm a quite unusual behavior is observed (sample series FS2406). For increasing $d_{CuNi}$ the critical temperature initially steeply drops towards zero at $d_{CuNi} \approx 4.0$ nm, i.e. the superconductivity is fully suppressed. After a further increase of $d_{CuNi}$ to a value of about 17.4 nm the superconducting state recovers. This is a reentrant behavior of the superconducting state predicted by theory [12].

In sample FS712, $d_{Nb}$ is further reduced to an average value about 6.3 nm. It is close to the critical value $d_{Nb}^{cr} \sim 6.2$ nm below which the superconductivity of Nb film is fully suppressed if the $Cu_{41}Ni_{59}$ alloy layer is sufficiently thick (see discussion below). As a result, upon increasing $d_{CuNi}$, the critical temperature $T_c$ steeply drops to zero and remains vanishing for any $d_{CuNi} \gtrsim 4$ nm.

In addition to the previous $T_c(d_{CuNi})$ dependencies, we also measured $T_c(d_{Nb})$ for a $Cu_{1-x}Ni_x$ layer of fixed thickness with a niobium layer of variable thickness on top (see Fig. 5b). For decreasing niobium thickness the critical temperature decreases. For not too low thickness of the niobium layer the decrease is relatively smooth and becomes steep in the range of very thin niobium layer thicknesses, *i.e.* 7-9 nm. The choice of niobium thicknesses in the range of the steep decrease is a key requirement to observe the deep oscillations of the critical temperature in the $T_c(d_{CuNi})$ measurements reported above. To get reentrant superconductivity, or even a multi-reentrant state, $d_{Nb}$ has to be very close to the critical thickness for which the critical temperature vanishes in Fig. 5b. The critical thickness, $d_{Nb}^{cr}$, can be estimated by fitting the theoretical curve to the $T_c(d_{Nb})$ measurements, as will be done in the next subsection.

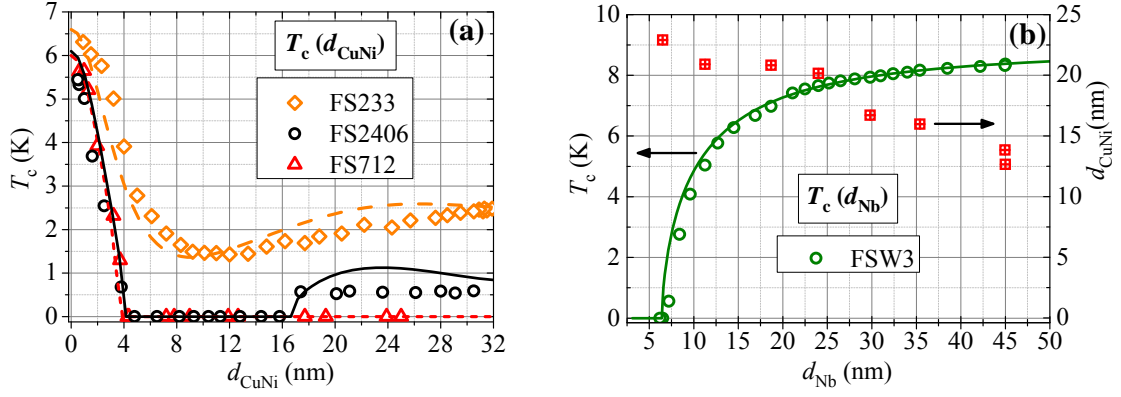

Fig. 5. a) Critical temperature $T_c$ of Nb/Cu$_{41}$Ni$_{59}$ bilayers as a function the thickness $d_{CuNi}$. For samples series FS233, FS2406, and FS712 it is $d_{Nb}$ = 7.5 nm 6.8 nm and 6.3 nm, respectively. b) Critical temperature $T_c$ of Nb/Cu$_{41}$Ni$_{59}$ bilayers as a function of $d_{Nb}$. For a detailed discussion see the text.

It should be noticed that the thickness of the CuNi layer in Fig. 5b is not completely constant, but increases from about 12.5 nm to 21 nm along the wedge of decreasing $d_{Nb}$. Most of this increase, however, takes place in the flat range of the $T_c(d_{Nb})$ curve where $d_{Nb}$ is relatively thick and, thus, less sensitive to changes of the thickness of the CuNi alloy layer. In the range where the $T_c(d_{Nb})$ curve steeply decreases towards zero, the thickness of the CuNi layer does nearly not change. Between $d_{Nb}$ = 18.6 nm ($T_c$ about 7 K) and $d_{Nb}$ = 7.5 nm ($T_c$ about 0.6 K) the CuNi thickness only slightly varies from 21 nm to about 22.5 nm. These variations more or less not affect the range of the steep decrease of the critical temperature, so that the extrapolation of the critical thickness should result in a reliable value.

### 3.2 Comparison of the Experimental Data with Theory

The characteristic feature of the proximity effect in bilayers of superconductors and ferromagnets is an oscillation of the superconducting pairing wave function on a wavelength scale $\lambda_{FM}$ during its decay into the ferromagnetic material. This wavelength scale is determined by the magnetic coherence length $\xi_F$. In the case of a clean ferromagnet ($l_F \gg \xi_{F0} = \hbar v_F/E_{ex}$, where $l_F$ and $v_F$ are the electron mean free path and the Fermi velocity in the F metal, respectively, $E_{ex}$ is the exchange splitting energy of a free-electron-like parabolic conduction band) it is $\lambda_{FM} = \lambda_{F0} = 2\pi\xi_{F0}$ [44,45]. In the dirty case ($l_F \ll \xi_{F0}$) it is $\lambda_{FM} = \lambda_{FD} = 2\pi\xi_{FD} = 2\pi(2\hbar D_F/E_{ex})^{1/2}$ [8,46], where $D_F = l_F v_F/3$ is the diffusion coefficient of electrons in the ferromagnetic material. On the other hand, the decay length of the pairing wave function is given by $l_F$ and $\xi_{FD}$ in the clean and dirty case, respectively [12,45,46].

The oscillatory behavior of the pairing wave function results in interference effects in bilayers of a superconducting and a ferromagnetic film, which can change periodically between the constructive and destructive case for increasing thickness of the F material. This results in a periodic modulation of the pairing function flux through the interface yielding $T_c$ oscillations as a function of $d_F$ [12,13,40].

In our previous experiments on S/F bilayers we found a consistent description of the data applying an extension of the theory given in Ref. [12] on the intermediate region $l_F \approx \xi_{F0}$, i.e. the crossover region between the dirty and the clean cases, represented by our samples. The expressions to calculate the superconducting $T_c$ for bilayers of a superconductor and a ferromagnet given in Ref. [12,14] were transformed for the use of physical parameters obtained from the experiments as mentioned in Ref. [40] and elaborated in detail in the Appendix of Ref. [14]. In the following we will apply these equations to calculate the critical temperature of our $Cu_{1-x}Ni_x$/Nb (i.e. F/S) bilayers.

As it was discussed above, from the vanishing of $T_c$, in Fig. 5b, the critical thickness at which superconductivity vanishes can be extrapolated by fitting the measurement by the corresponding theoretical curve. This procedure results in $d_{Nb}^{cr} \approx 6.2$ nm. The theoretical curve was calculated for the CuNi layer thickness $d_F/\xi_F = 2.0$ ($d_F \approx 22$ nm). Although the thickest and almost constant part of the CuNi film is in the range of the steep decrease of the critical temperature (as discussed above), it is not of physically infinite thickness as required by the theory. This can be seen e.g. from Fig. 5a in which the critical temperature still changes above $d_{CuNi}$ between 21 nm and 22.5 nm. Nevertheless, the value obtained for the critical thickness seems us to be reliable enough to impose a constraint on the parameters $N_F v_F/N_S v_S$ and $T_F$ using the expression [39,47],

$$d_{Nb}^{cr} = \xi_S \sqrt{2\gamma} \arctan\left\{\frac{\pi}{\sqrt{2\gamma}}\left(\frac{\xi_{BCS}}{\xi_S}\right)\frac{N_F v_F / N_S v_S}{1 + 2/T_F}\right\}, \quad (1)$$

as follows

$$\frac{N_F v_F}{N_S v_{SF}}\frac{1}{1 + 2/T_F} \approx 0.052, \quad (2)$$

where $\xi_S = 6.0$ nm and $\xi_{BCS} = 42$ nm [48] have been used ($\xi_{BCS}$ is the Bardeen-Cooper-Schriefer coherence length, see e.g. [14], see notations of the other parameters below), $\gamma \approx 1.781$ is the Euler constant. Now, $N_F v_F/N_S v_S$, as well as $T_F$, can be varied, however, around the condition fixed by Eq. (2). With this reduction of the parameters the problem of consistent fitting the curves in Figs. 5a and 5b becomes possible.

The five physical parameters (reduced to three by the constraint of Eq. (2)) which enter the theory are: $\xi_S$, the superconducting coherence length in a superconducting metal as e.g. defined in Eq. 1 of [14]; $\xi_{F0}$, the coherence length for Cooper pairs in a ferromagnetic metal; $l_F$, the mean free path of conduction electrons in a ferromagnet; $N_F v_F/N_S v_S$, the ratio of Sharvin conductance's at the S/F interface, and $T_F$, the interface transparency parameter. A first guess for the value of the superconducting coherence length $\xi_S$ is obtained from upper critical field measurements, yielding a range from 6.2 nm to 6.7 nm [14].

For the calculation of the curves in Fig. 5 the following parameters were used for sample series FS233, FS2406, and FS712. Critical temperatures for a niobium thin film in the absence of a ferromagnetic copper-nickel layer $T_{c0,Nb}(d_{CuNi} = 0\text{nm}) = 6.6$ K, 6.1 K, and 6.0 K, respectively; $\xi_S = 6.0$ nm, 6.0 nm, and, 6.0 nm; $N_F v_F/N_S v_S = 0.19$, 0.20, and 0.20; $T_F = 0.81$, 0.88, and 0.82; $l_F/\xi_{F0} = 0.6$, 0.6, and 0.7; $\xi_{F0} = 11.6$ nm, 10.4 nm, and 10.8 nm. The curve in Fig. 5b was calculated using $T_{c0,Nb} = 8.9$ K, $\xi_S = 6.0$ nm, $N_F v_F/N_S v_S = 0.22$, $T_F = 1.0$, $l_F/\xi_{F0} = 0.9$, $d_F/\xi_{F0} = 2.0$, where $\xi_{F0} = 11.0$ nm.

The calculated curves in Fig. 5a seem not perfectly match the measured points, especially in the range of the recovered superconductivity ($d_{CuNi} > 16$ nm). Our explanation is that the recovered superconductivity is very sensitive to any perturbation of the interference of the pairing function which is the underlying reason of the oscillating or reentrant behavior of the superconducting $T_c$. The interface roughness, lateral inhomogeneities of the Ni concentration in the CuNi alloy, not taken into account in the modeling, can disturb the interference conditions thus preventing full recovery of the superconductivity presented by our calculated curves.

In principle, the superconducting coherence length $\xi_S$ depends on the electron mean free path, $l_s$, and the critical temperature, $T_{c0}$, of the niobium layer, while $\xi_{BCS}$ depends only on $T_{c0}$ of the niobium layer. The $l_s$ as well as $T_{c0}$ vary with the thickness of the layer. As we already discussed in Ref. [14] our experience shows that neglecting such kind of corrections using data for a free standing niobium film does not affect the physical considerations (compare also our discussion of Fig. 6 in Ref. [40], in which corrections concerning the critical temperature were made).

As the electron mean free path of the ferromagnetic material in our calculations, $l_F \approx 6.2\text{-}7.6$ nm, is of the order of the coherence length, $\xi_{F0} = 10.4\text{-}11.6$ nm, our samples are neither in the clean nor in the dirty limit. This means that the extension of the dirty-case theory towards the clean case, applied on the intermediate region, has been used to describe the experiment, as mentioned above.

Our analysis of the data results in parameters which are similar to the case of S/F bilayers, except the transparency, $T_F$, the ratio $l_F/\xi_{F0}$, and the electron mean free path $l_F$ in the F-metal. The values of $T_F$ for the F/S bilayers are about 1.4 times higher than for the S/F bilayers case, the ratio $l_F/\xi_{F0}$ is about a factor of 2 smaller, and $l_F$ is a factor of 1.6-1.9 smaller. Since the observed effects are similar for F/S and S/F bilayers probably the higher transparency compensates the smaller value of $l_F$, which represents the decay length of the pairing wave function in the F metal in the clean case. The minimum of the critical temperature in Fig. 5a in series FS233 is situated at about $d_{CuNi} = 10$ nm. This is at a somewhat higher value compared to our results for S/F bilayers [14], where this value is around 7-8 nm. This seems to indicate a decrease of the exchange energy. The reason is that the position of the minimum should occur around a fixed ratio of $d_F/\xi_{F0}$, according to the theory [12]. Thus, an increase of $d_F$ of the position of the minimum of $T_c$ indicates an increase of $\xi_{F0}$. Since $\xi_{F0} = \hbar v_F/E_{ex}$, this yields a decrease of the exchange energy as mentioned above. Therefore, the reduced value of $l_F/\xi_{F0}$ could also partly be caused by an increase of $\xi_{F0}$. In this case the reduction of $l_F$ would be not as strong as discussed above.

**Conclusion**

In the present paper the critical temperature of F/S bilayers is investigated as a function of the thickness of the ferromagnetic material. A deep oscillation of the critical temperature, a reentrant superconducting state and a steep transition to the normal state were observed, if the thickness of the flat niobium layer is gradually decreased. This is just the behavior predicted by the theory for ferromagnet-superconductor bilayers. Our results for F/S bilayers have been achieved for different thicknesses of the functional layers compared with the S/F bilayers.

Since in the present experiments (F/S sequence) the superconducting material is grown on the ferromagnetic layer, contrary to the S/F systems where the opposite sequence was the case, we succeeded in a key step to the realization of a Ferromagnet/Superconductor/Ferromagnet trilayer, representing the core structure of the superconducting spin valve. Now, both technological regimes can be combined to provide a coherent reentrant behavior of the both bilayers in an F/S/F trilayer (see Fig. 1) to get a spin-switch effect with a critical temperature shift in the Kelvin range.

**Acknowledgments**

The authors are grateful to S. Heidemeyer, B. Knoblich and W. Reiber for assistance in the TEM sample preparation, and A. S. Prokofyev for assistance in a part of the low temperature measurements. The work was supported by the Deutsche Forschungsgemeinschaft (DFG) under the grant No GZ: HO 955/6-1, and in part by Russian Fund for Basic Research (RFBR) under the grant No 09-02-12260-ofi_m (L.R.T.).